\documentclass[varenna]{cimento}
\usepackage{graphicx}
\usepackage{float}

\begin{document}

  \bibliographystyle{unsrt}
  \title {Catastrophic Cascade of Failures in Interdependent Networks}

  \author{S. Havlin}
    \institute{Minerva Center and Department of Physics, Bar-Ilan University, 52900 Ramat-Gan, Israel}

  \author{N. A. M. Ara\'ujo}
    \institute{Computational Physics for Engineering Materials, IfB, ETH Z\"{u}rich, Schafmattstr. 6, 8093 Z\"{u}rich, Switzerland} 

  \author{S. V. Buldyrev}
    \institute{Department of Physics, Yeshiva University, 500 West 185th Street, New York 10033 USA}
    \institute{Center for Polymer Studies and Department of Physics, Boston University, Boston Massachusetts 02215, USA}  

  \author{C. S. Dias}
    \institute{GCEP-Centro de F\'isica da Universidade do Minho, 4710-057 Braga, Portugal}

  \author{R. Parshani}
    \institute{Minerva Center and Department of Physics, Bar-Ilan University, 52900 Ramat-Gan, Israel}

  \author{G. Paul}
    \institute{Center for Polymer Studies and Department of Physics, Boston University, Boston Massachusetts 02215, USA}  

  \author{H. E. Stanley}
    \institute{Center for Polymer Studies and Department of Physics, Boston University, Boston Massachusetts 02215, USA}  

\maketitle

\begin{abstract}
 Modern network-like systems are usually coupled in such a way that failures in one network can affect the entire system.
 In infrastructures, biology, sociology, and economy, systems are interconnected and events taking place in one system can propagate to any other coupled system.
 Recent studies on such coupled systems show that the coupling increases their vulnerability to random failure.
 Properties for interdependent networks differ significantly from those of single-network systems.
 In this article, these results are reviewed and the main properties discussed.

\end{abstract}

\section{Introduction}

  The last decade witnessed an intensive study of complex networks \cite{Albert02,Dorogovstev02,Newman03,Newman10,Dorogovstev08,Cohen10} boosted by several real-world data revealing complex structures in the topology of their network like, Internet, airport connections, and power grids \cite{Clauset09}.
  Recently, special emphasis has been focused on the robustness of such systems to random failures or malicious attacks \cite{Albert00, Cohen01, Albert04, Moreira09, Holme02, Schneider10, Herrmann10}.
  Most of these works have been focused on single, isolated networks where no interaction with other networks is considered, i.e., the behavior of the system is independent of any other, coupled with it.
  Such conditions rarely occur in nature nor in technology.
  Typically, systems are interconnected and events taking place in one are likely to affect the others.
  Only recently \cite{Buldyrev10,Parshani10}, the effect of coupled networks has been considered, where a failure of one node in a network may lead to a cascade of failures in the entire system.
  In this manuscript we review these results, obtained through analytical and numerical approaches, based on percolation principles, which are rather surprising.

  \begin{figure}[t]
  \begin{center}
    \includegraphics[width=10.0cm]{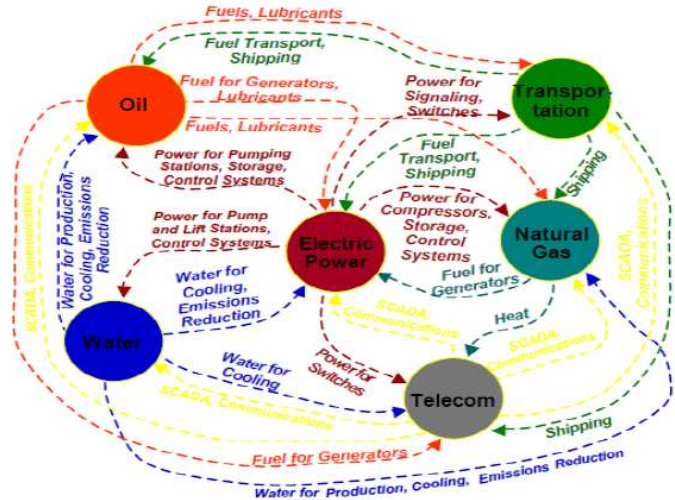}\\
  \end{center}
    \caption{
      Interdependency of infrastructures (After Peerenboom {\it et al.} \cite{Peerenboom01}).
      \label{fig:infrastructures}
    }
  \end{figure}

\begin{figure}[t]
  \begin{center}
    \includegraphics[width=13.0cm]{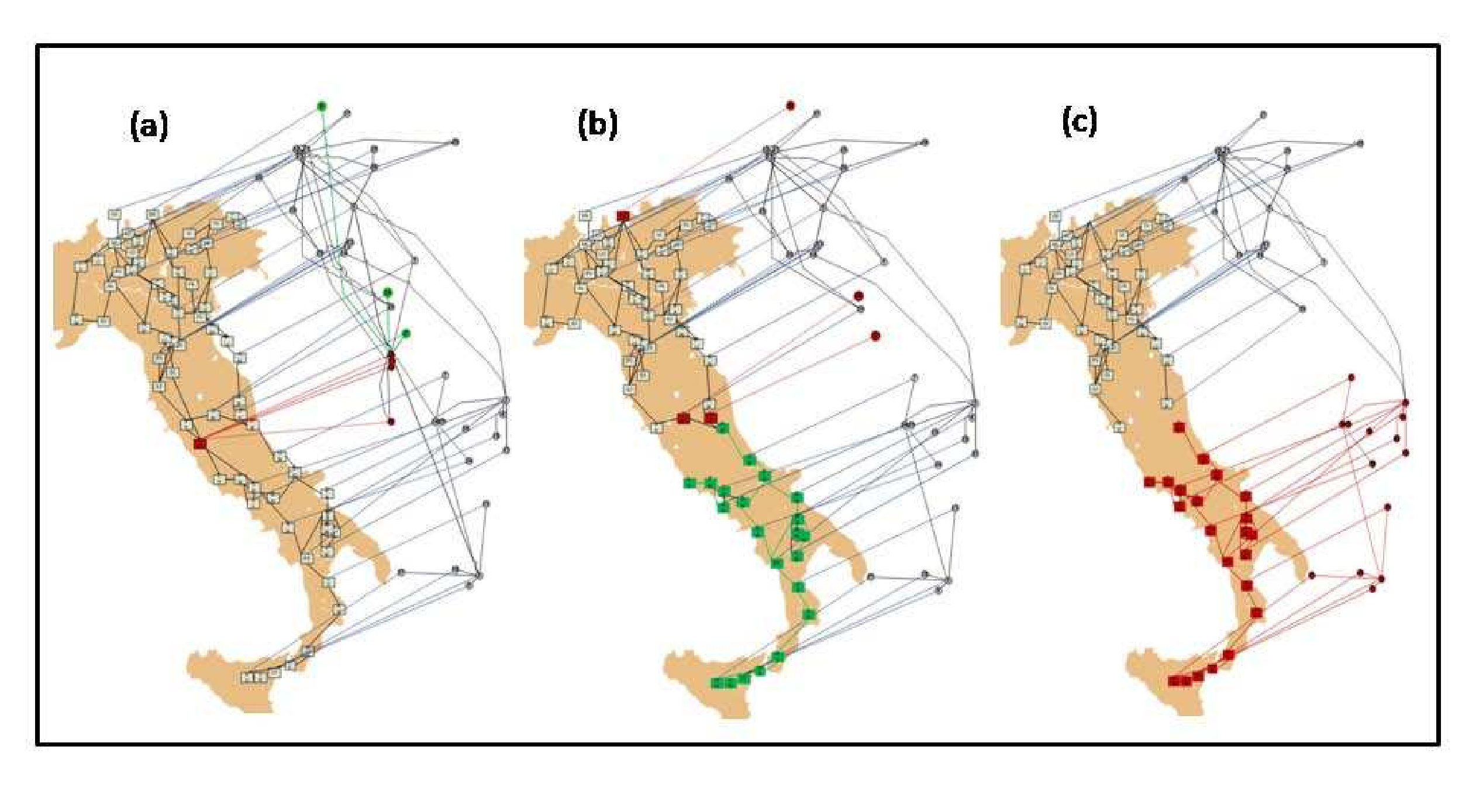}\\
  \end{center}
    \caption{
      Cartoon of a typical cascade obtained by implementing the described model on the real coupled system in Italy. 
      Over the map is the network of the Italian power network and, slightly shifted to the top, is the communication network. Every server was considered to be connected to the geographically nearest power station.
      (After Buldyrev {\it et al.} \cite{Buldyrev10})
      \label{fig:Italy}
    }
  \end{figure}

  In Fig.~\ref{fig:infrastructures} is a scheme, from Peerenboom {\it et al.} \cite{Peerenboom01}, showing the interdependence of the relevant infrastructures for daily life, namely, oil, transportation, electric power, natural gas, water, and telecommunications.
  Not only all infrastructures depend on the electrical-power network, as one would expect, but also electrical power depends on the others, i.e., a bidirectional coupling exists between all networks.
  This strong coupling can lead to catastrophic effects when, for some reason, a failure occurs in one of the networks.
  In September $28^{\mbox{\scriptsize th}}$, $2003$, Italy was affected by a country-wide blackout \cite{Rosato08}, which was understood due to the coupling between the electrical power and communication networks.
  In Fig.~\ref{fig:Italy} is the map of Italy with the electrical power network on top.
  Each node is a power station and the edges represent the connection between stations.
  Slightly shifted to the top, is a scheme of the communication network that controls the power distribution, where the nodes are servers and the edges are links between them.
  In both cases, nodes are positioned based on their real geographical coordinates and communication servers are connected to their geographically nearest power station.
  When, for some reason, a failure occurs in one power station (red node in Fig.~\ref{fig:Italy}.a), the node is removed from the network, and consequently four servers are turned off due to the lack of power supply.
  When these servers go down, three other servers (in green) become inactive since they are disconnected from the giant communication cluster.
  Then, a sequence of events takes place, in a cascade fashion, where the power stations connected to these servers are shut down (see Fig.~\ref{fig:Italy}.b) and a set of other power stations (in green) become disconnected from the power grid giant component and, therefore, become inactive (see Fig.~\ref{fig:Italy}.c).
  This example shows how a fail in a single power station can lead to a cascade of events ending in a blackout spanning over more than half of the system.

  Not only in infrastructures one finds interdependent networks.
  From economy to biology there exist many examples of coupled systems.
  The network of banks is interdependent to the network of insurance companies, as well as to the network of firms in different fields.
  These interconnections played a major role in the recent financial crisis.
  The human body networks are also interdependent, e.g., the cardio-vascular, the respiratory, and the nervous systems depend on each other in order to function.
  Examples of interdependent networks can also be found in social sciences.
  Therefore, the study of the properties of coupled networks is a matter of paramount importance in multidisciplinary science. 
  In this article, we review the recent analytical and numerical results on their percolation properties.
  We start by recalling the main features of single networks, in Sec.~\ref{sec:SN}. 
  In Secs.~\ref{sec:IDN}~and~\ref{sec:PDN} the case of fully and partially interdependent networks are discussed. 
  We present final remarks in Sec.~\ref{sec:FR}.

\section{Single Network Robustness}\label{sec:SN}

  For more than a decade the critical properties of isolated (single) networks have been studied extensively, see e.g., \cite{Dorogovstev08,Cohen10}.
  One relevant question is related with percolation, i.e., the emergence of a giant cluster when nodes (or links) are sequentially added to the network \cite{Stauffer94, Bunde96, Erdos59, Erdos60, Bollobas85}.
  Due to symmetry reasons, for a single network, the problem can also be formulated in the inverse way.
  Let us consider an initial configuration of a network made of nodes and links connecting them.
  How the fraction of nodes in the giant cluster (largest one) is changed when a fraction $1-p$ of nodes (or links) is removed? The fraction of the giant component is called the order parameter in the language of critical phenomena \cite{Stauffer94, Bunde96}.
  
  \begin{figure}[t]
  \begin{center}
    \includegraphics[width=10.0cm]{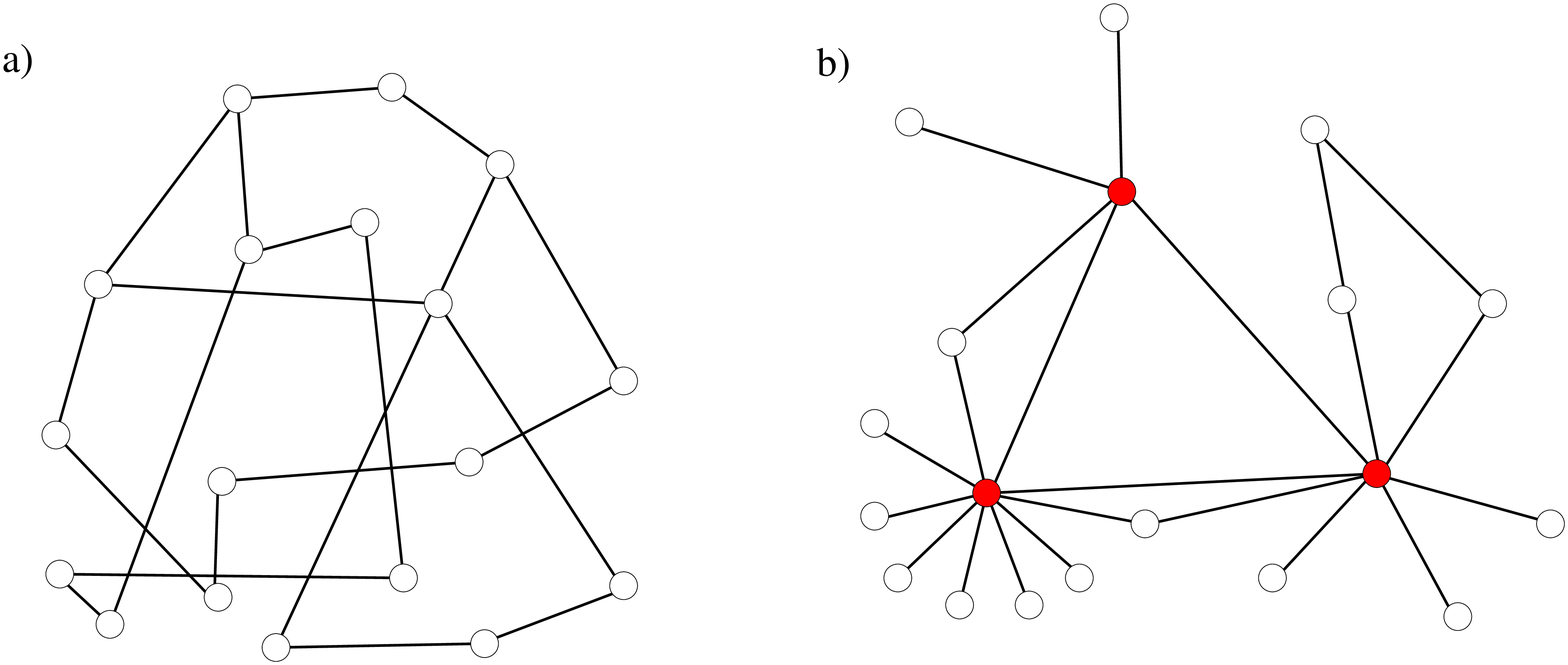}\\
  \end{center}
    \caption{
      Single networks. a) Classical Erd\H{o}s-R\'enyi model and b) Barab\'asi-Albert model.
      \label{fig:single}
    }
  \end{figure}

  Several models of networks have been proposed.
  Their description and critical properties can be found elsewhere \cite{Dorogovstev08,Cohen10}.
  In this article we focus on two types of model networks: a random graph (Erd\H{o}s-R\'enyi) \cite{Erdos59, Erdos60, Bollobas85} and a scale-free network (Barab\'asi-Albert) \cite{Barabasi99}.
  The Erd\H{o}s-R\'enyi (ER) network is a random graph obtained by randomly distributing $M$ links between $N$ nodes, being a statistical ensemble with equal probability for any generated configuration. 
  The scheme in Fig.~\ref{fig:single}.a is one possible configuration.
  On the other hand, the Barab\'asi-Albert (BA) is a network which was grown under the preferential attachment rule, i.e., at each iteration a new node is added to the network and connected to $m$ already existing nodes with a probability of linking to a certain node proportional to the actual degree (number of links) of that node.
  A typical configuration obtained with this model is shown in Fig.~\ref{fig:single}.b.

  These two models of generating networks lead to different topologies and statistical properties.
  For the ER network, since links are distributed in an uncorrelated way, the degree distribution is Poissonian, i.e., the frequency of nodes with $k$ links is \cite{Erdos59},

    \begin{equation}
      P(k)=\exp{\left(-\lambda\right)}\frac{\lambda^k}{k!} \ \ ,
      \label{eq:erdos}
    \end{equation}

  \noindent where $\lambda$ is the average degree, $\lambda=<k>$, of the entire graph.
  While in the BA network, the preferential attachment rule develops correlations in the network and a scale-free degree distribution is obtained, which in a general way is given by \cite{Barabasi99},

    \begin{equation}
      P(k)=\left\{ \begin{array}{cl}
           ck^{-\gamma} &\mbox{$m\leq k<K$} \\
           0 &\mbox{ otherwise}
           \end{array} \right. \ \ ,
          \label{eq:barabasi}
    \end{equation} 

  \noindent where $\gamma$ is the degree exponent and $K$ an upper limit due to the system finite size.
  The power-law degree distribution favors the existence of highly connected nodes when compared with the Poisson distribution.
  This feature is, in fact, responsible for relevant differences in the critical properties of networks obtained with these two models \cite{Dorogovstev08, Cohen02}.

  To study the percolation properties of a network, a fraction 1-p of nodes are randomly removed, and the behavior of the giant component is studied.
  For an ER network, a critical fraction of nodes $p_c$ can be defined such that, in the thermodynamic limit, the fraction of sites in the giant component goes to zero when more than $1-p_c$ fraction of nodes are removed.
  A transition then occurs from a percolative ordered state, characterized by the existence of a nonzero giant component to a nonpercolative one, where the size of the giant component is zero.
  The order parameter of this phase transition is then the fraction of sites in the giant cluster, $P_\infty$, where for $p>p_c$,

  \begin{equation}\label{eq:order.param}
    P_\infty\sim(p-p_c)^\beta \ \ ,
  \end{equation}
  \noindent and $\beta$ is a critical exponent.
  The transition is second order and the percolation threshold for ER networks is given by $p_c=1/<k>$ \cite{Bollobas85}.

  The power-law nature of the degree distribution in a scale-free graph leads to richer percolation properties in this type of networks.
  In fact, such properties are dependent on the value of the degree exponent $\gamma$.
  For $2\leq\gamma\leq3$, the giant component only vanish for $p_c=0$ and no percolation transition occurs \cite{Cohen00}.
  However, for values of $\gamma$ above 3, alike ER a second-order transition is found but with different critical exponents \cite{Dorogovstev08, Cohen02}.
  Yet, the network is very robust to random failure which explains, for example, the stability of the Internet to random failures and the high longevity of viruses in the Internet, despite the large number of anti-virus softwares in the market.

  For both types of networks, when the percolation transition occurs, it is of second-order nature, meaning that a smooth decrease of the order parameter with the increasing fraction of removed nodes is observed.
  Besides, this fraction, at the critical point, is rather large ($p_c$ is low) being a sign of robustness in the system.
  As discussed below, this is not the case, when interdependent systems are considered.
  Recently, several studies have been published, discussing how to change the stochastic rule of percolation to obtain an abrupt transition \cite{Achlioptas09, Ziff09, Radicchi09, Radicchi10, Araujo10, DaCosta10} in a single network.
  When coupled interdependent systems are considered the order of the transition becomes naturally of first order nature \cite{Buldyrev10,Parshani10}.

\section{Interdependent Networks Robustness}\label{sec:IDN}

  Despite the technological and fundamental relevance of coupled networks, only recently they have been considered and their percolation properties have been studied \cite{Buldyrev10,Parshani10}.
  The results of Buldyrev {\it et al.} \cite{Buldyrev10} disclose an emergency of novel percolation properties under coupling not predicted from the behavior of single networks.

  To account for the coupling, let us consider two different networks, hereafter referred as $A$ and $B$.
  Every $A$-node depends on a $B$-node, and vice versa, i.e., a bidirectional, one-to-one coupling exists such that if node $A_i$ depends on node $B_i$ then node $B_i$ depends on node $A_i$.
  The dependency is such that coupled nodes are only active if both are connected to the giant component of their network.
  Network $A$ and network $B$ have degree distributions $P_A(k)$ and $P_B(k)$ respectively. 
  The dependency links between the networks are achieved by randomly connecting $(A_i,B_i)$ pairs of nodes in both networks under the constraint of having only one inter-network link per node.

  Percolation properties are studied by randomly removing nodes in the network $A$, mimicking a failure or malicious attack.
  When a node $A_i$ is removed, its $A$-links and the coupled node $B_i$ are also removed.
  As discussed above, in the context of the 2003 blackout in Italy, removing the node $A_i$ can ignite failures in other nodes.
  All $A$-nodes which become disconnected from the giant cluster through $A_i$ become inactive and are removed together with their correspondent $B$-nodes.
  Analogously, all $B$-nodes bridged to the giant component through node $B_i$ are also removed.
  The same procedure is, recurrently, followed for all removed nodes and their counterparts in the other network leading to cascading failures between the two networks.
  This procedure reveals, as in the case of single networks, discussed in Sec.~\ref{sec:SN}, that when a fraction of nodes $1-p$ is removed, a percolation transition occurs at a certain threshold, $p=p_c$.
  Only for values of $p$ above this threshold a giant mutually connected cluster exists.
  Below it, the entire system becomes completely fragmented.

  \begin{figure}[t]
    \begin{center}
      \includegraphics[width=8.0cm]{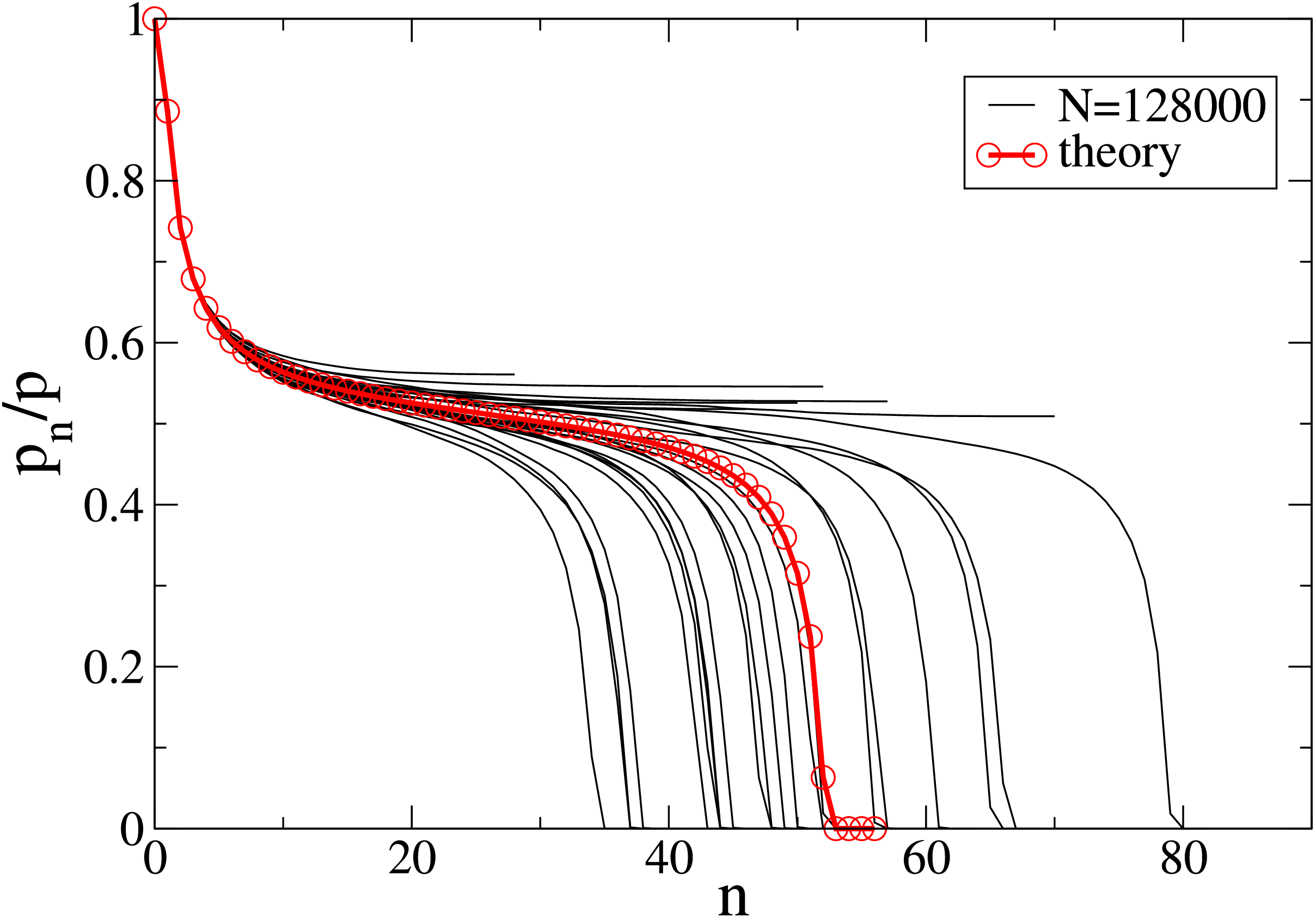}\\
    \end{center}
      \caption{Fraction of nodes in the giant component ($p_n/p$) after $n$ stages of cascading failures for different realizations of two coupled ER networks with the same degree distribution and $128000$ nodes.
        Initial removal is just below the percolation threshold with $p=2.4554/<k>=p_c$. (After Buldyrev {\it et al.} \cite{Buldyrev10}).
        \label{fig:cascade}
      }
  \end{figure}

  For two coupled ER networks the problem can be solved explicitly using generating functions \cite{Buldyrev10,Newman02,Shao08,Shao09}.
  When the two networks have the same degree, i.e., $<k>_A=<k>_B=<k>$, the value of $p_c$ is,
  \begin{equation}\label{eq:pc.ER}
    p_c=\frac{1}{2<k>f(1-f)} \ \ ,
  \end{equation}
  \noindent where $f=exp((f-1)/2f)$, with the root $f\approx0.28467$, giving $p_c\approx2.4554/<k>$.
  This threshold is much larger than the one for single networks, where $p_c=1/<k>$ (see Sec.~\ref{sec:SN} for details), revealing a significant increase in vulnerability due to the coupling.
  Moreover, if the fraction of nodes in the giant component ($\mu_\infty$) is analyzed, as a function of $p$, unlike the single network case, where it continuously vanishes when approaching $p_c$, a nonzero finite value is obtained at $p_c$.
  The cascade of failures leads to a first-order transition, characterized by a jump in the order parameter (fraction of nodes in the mutually connected giant component) at $p_c$.

\begin{figure}[t]
    \begin{center}
      \includegraphics[width=8.0cm]{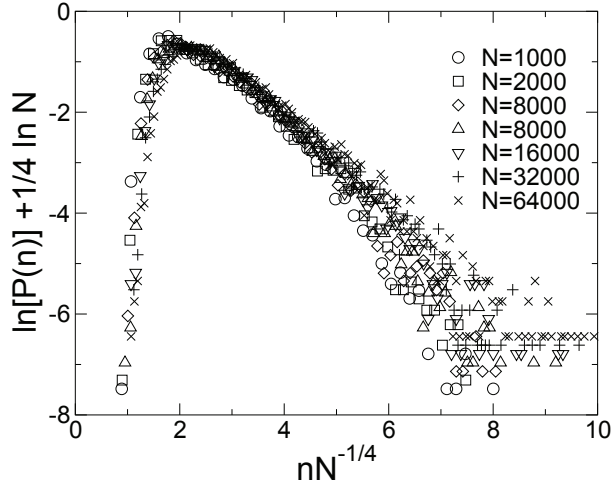}\\
    \end{center}
    \caption{
      Scaled distribution of the number of stages in the cascade failures for two Erd\H{o}s-R\'enyi graphs, with the same degree distribution, at criticality, for different values of $N$. (After Buldyrev {\it et al.} \cite{Buldyrev10}).
      \label{fig:distribution}
    }
  \end{figure}

  The fraction of the giant component, $p_n/p$, after $n$ stages of the cascades for different realizations of ER networks with $128000$ nodes, near $p_c$, is shown in Fig.~\ref{fig:cascade}.
  For simplicity, the same degree distribution for both networks have been considered.
  As predicted by the theory, after a plateau of many cascading steps, the fraction of the giant component suddenly drops to zero.
  Due to the system finite size, for simulations, the system converges either to a mutual giant component or to a complete fragmentation of the system.
  At $p_c$, the average number number of stages in the cascade scales with $N^{1/4}$ (see Fig.~\ref{fig:distribution}).
  
  \begin{figure}[t]
    \begin{center}
      \includegraphics[width=8.0cm]{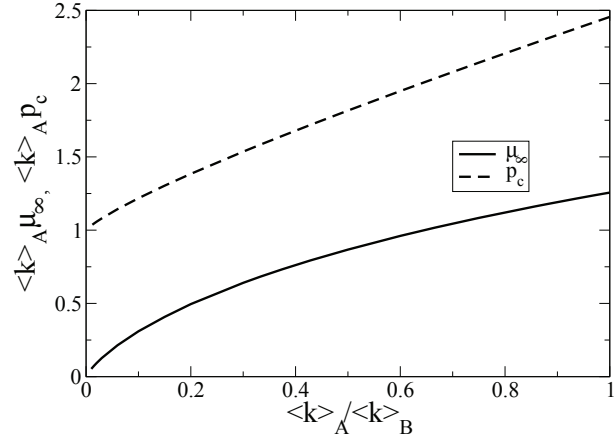}\\
    \end{center}
    \caption{
      For two coupled Erd\H{o}s-R\'enyi networks, the threshold $p_c$ and the fraction of nodes in the mutual giant cluster at the transition, $\mu_\infty$, are plotted  as a function of the ratio between the average degrees of networks A and B, $<k>_A/<k>_B$.
      \label{fig:ER_ka_kb}
    }
  \end{figure}

  In Fig.~\ref{fig:ER_ka_kb} we plot the threshold, $p_c$, and the fraction of nodes in the mutual giant component, $\mu_\infty$, for different ratios between the average degrees of the networks, $<k>_A/<k>_B$.
  When the average degree is the same for both networks, $<k>_A=<k>_B$, as discussed above, the threshold is given by eq.~(\ref{eq:pc.ER}) and $\mu_\infty$ at $p_c$ is nonzero, i.e., the coupled system is more vulnerable than the single network case and the percolation transition is first order.
  As the ratio between the average degree of the networks decreases, both the threshold and the jump at the transition diminish.
  Therefore, the smaller the ratio the more resilient is the system to failures.
  In the limit where the ratio approach zero, the single-network features are recovered, i.e., the transition becomes of second-order nature, with $p_c=1/<k>$.

  \begin{figure}
    \begin{center}
      \includegraphics[width=8.0cm]{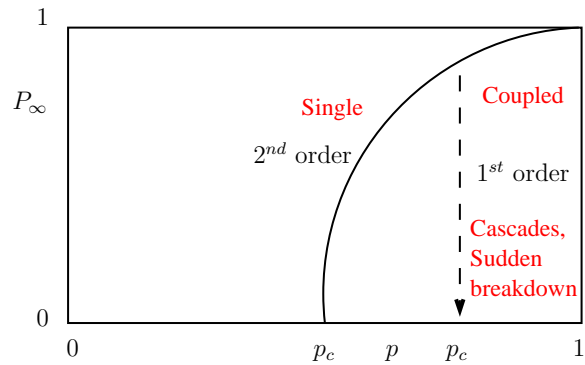}\\
    \end{center}
    \caption{
      The order parameter (fraction of sites in the giant component) as a function of the fraction of left nodes in single and coupled networks.
      For single networks, above the percolation threshold a smooth increase of the order parameter is observed, with a critical exponent $\beta$.
      However, for strongly coupled networks, an abrupt transition is observed, characteristic of first-order transitions, due to the cascade phenomenon discussed in the text.
      \label{fig:large_clust_single}
    }
  \end{figure}

  When coupled scale-free networks are considered, a percolation transition, at nonzero $p_c$, is obtained even for values of the degree exponent $2<\gamma\leq3$.
  This is in contrast to the single network case, where a giant cluster is observed for any positive fraction $p$ of nodes (see Sec.~\ref{sec:SN}).
  Analogously to ER networks, the coupling between scale-free networks significantly increases the vulnerability of the system, with a larger $p_c$ compared to the case of a single network.
  Since hubs can have a low-degree counterpart node, their vulnerability evinces with the coupling.
  In contrast to single networks, the broader the degree distribution the larger is $p_c$, i.e., the smaller the fraction of nodes that needs to be removed to fully fragment the system.
 
  In general, the coupling between networks, increase the vulnerability of the system due to the cascade of failures that can be activated by removing a small fraction of nodes.
  Findings are summarized in Fig.~\ref{fig:large_clust_single}.
  While for a single network, a second-order percolation transition is observed where the order parameter vanishes smoothly at criticality, for coupled systems, the transition occurs for larger $p_c$ (lower fraction of removed nodes), and is rather abrupt, characterized by a discontinuity in the order parameter. 
  In the next section, the case of partially dependent networks is reviewed.

\section{Partially Dependent Networks}\label{sec:PDN}

  \begin{figure}[t]
    \begin{center}
      \includegraphics[width=8.0cm]{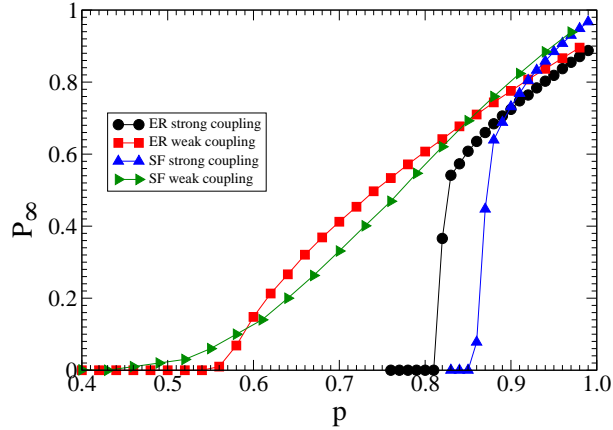}\\
    \end{center}
    \caption{
      Order parameter $P_\infty$ as a function of the fraction of nodes left $p$ for Erd\H{o}s-R\'enyi and scale-free network ($\gamma=2.7$) with strong and weak coupling.
      Both systems contain $5\times 10^4$ nodes.
      For both network types, first-order transitions occurs for strong coupling in contrast to second order transition in weak coupling. (After Parshani {\it et al.} \cite{Parshani10}). 
      \label{fig:coupling}
    }
  \end{figure}

  Frequently, in real-world systems not all the nodes in one network are interdependent on the other network and vice versa, via bidirectional links.
  Instead, some nodes may be autonomous and independent on nodes from the other networks.
  For example, in the communication/power-grid system, a fraction of servers may be protected by emergency power supplies which are activated when the local power station is shut down.
  Parshani {\it et al.} \cite{Parshani10} have studied the behavior of partially interdependent networks under random failure and found quantitatively how reducing the coupling improves the robustness of the system against random attack and how percolation transition changes from first to second order.

  The model introduced in Sec.~\ref{sec:IDN}, can now be generalized to account for partially coupling between networks.
  Two coupled networks are then considered, $A$ and $B$, with degree distributions $P_A(k)$ and $P_B(k)$, respectively.
  In network $A$, a fraction of nodes $q_A$ depends on nodes from network $B$.
  Likewise, in network $B$, a fraction of nodes $q_B$ depends on nodes in $A$.
  In the limit of $q_A=q_B=1$ the fully interdependent system, discussed above, is recovered.
  
  \begin{figure}
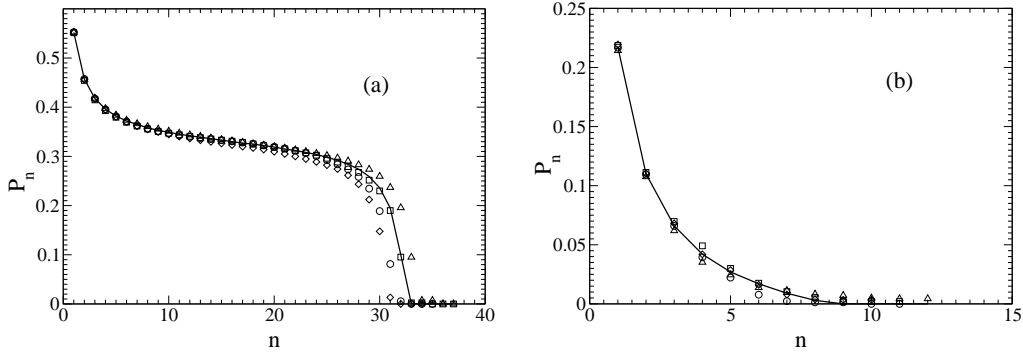

    \begin{center}
    \begin{tabular}{cc}
      \includegraphics[width=6.5cm]{./9a.eps} & \includegraphics[width=6.5cm]{./9b.eps}\\
    \end{tabular}
    \end{center}
    \caption{
      The giant component as a function of the number of cascade failures in two coupled Erd\H{o}s-R\'enyi networks, with $N_A=N_B=8\times10^5$ and $<k>_A=<k>_B=2.5$, for two different coupling strengths. (a) $p=0.7455$, $q_A=0.7$, and $q_B=0.6$ (first order). (b) $p=0.605$, $q_A= 0.2$, and $q_B=0.75$ (second order). Solid lines correspond to results obtained from theory based on generating functions. (After Parshani {\it et al.} \cite{Parshani10}).
      \label{fig:strong_weak_coupling}
    }
  \end{figure}

  Following the procedure described before, $1-p$ nodes are randomly removed and the percolation properties of the system studied.
  If a removed node has a dependent in the other network, this node is also removed.
  All nodes linked to the correspondent giant component solely through the removed nodes are also considered inactive.
  In Fig.~\ref{fig:coupling} we show the curves of the fraction of sites belonging to the mutually connected giant component, $P_\infty$, as a function of $p$.
  ER and scale-free networks (with $\gamma=2.7$) have been considered, with strong ($q_A=q_B=q=0.8$) and weak ($q=0.1$) coupling.
  For both systems in the strong coupling case, the robustness of the coupled networks systems is similar to that observed in the limit of $q=1$ (see Sec.~\ref{sec:IDN}).
  Reducing the coupling leads to a second order phase transition similar to single networks (the case of $q=0$). Increasing the coupling leads to a percolation transition at larger $p_c$ and to a change from second order, under weak coupling, to first order for strong coupling.

  The giant component as a function of the number of iterations in the cascade of failures, close to the transition point, for a strong and weak coupled system is shown in Fig.~\ref{fig:strong_weak_coupling}.
  Two coupled ER networks have been considered, with $N_A=N_B=8\times10^5$ and the same average degree of nodes, with strong and weak coupling.
  In the strong coupling regime, a plateau is obtained followed by an abrupt decrease of the order parameter, similar to the case $q=1$ of the a first-order transition.
  While, in the weak coupling regime, the order parameter smoothly vanish with the number of iterations of the cascades.

\begin{figure}
    \begin{center}
      \includegraphics[width=8.0cm]{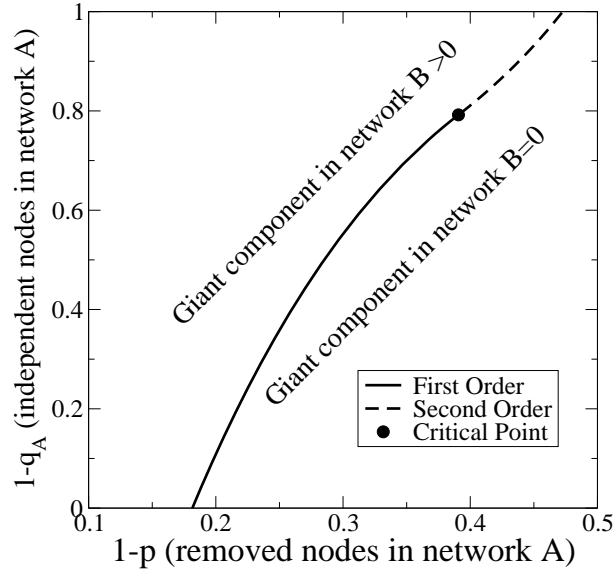}\\
    \end{center}
    \caption{
      Diagram for percolation in two coupled networks in the two-parameter space, namely, the fraction of independent nodes in one network and the fraction of removed nodes in the same network.
      For weak coupling (large fraction of independent nodes) the transition is continuous (dashed line).
      Below a certain fraction of independent nodes, due to the coupling strength, the transition becomes discontinuous (first order).
      The vulnerability of the system increases with increasing in the coupling strength. (After Parshani {\it et al.} \cite{Parshani10}).
      \label{fig:diagram}
    }
  \end{figure}

  Results for partially dependent networks are summarized in the two-parameter phase diagram of Fig.~\ref{fig:diagram}.
  In the horizontal axis is the fraction of removed nodes in network $A$, $1-p$, while in the vertical one is the fraction of independent nodes in the same network, $1-q_A$. The value of $q_B$ is chosen to be $q_B=1$.
  Two different regimes can be identified in the diagram.
  In the right side of the diagram (below the curve), no finite giant component exists in network $B$ at $p_c$. That is, the system is below $p_c$.
  In the left side, a giant component exists in the system.
  This two different regimes are separated by a phase transition line.
  When one crosses this line, a percolation transition is observed.
  The solid line corresponds to a first-order transition, characterized by a jump in the order parameter at the transition point.
  The dashed line stands for a second-order transition, where a continuous change of the order parameter is obtained.
  For a fully coupled system, $1-q_A=0$, only a small fraction of nodes needs to be removed to observe a fragmentation of the whole system in a discontinuous way.
  As one increases the fraction of independent nodes in the system, the fraction of removed nodes at the transition increases, meaning that the vulnerability of the system increases.
  This first-order line of the transition ends at a critical point above which the transition is no longer abrupt, being smooth and of second-order nature.
  
\section{Final remarks}\label{sec:FR}

  In this review, the recent work by Buldyrev {\it et al.} \cite{Buldyrev10} and Parshani {\it et  al.} \cite{Parshani10} on the percolation properties of interdependent networks have been revisited. The review is based on a Lecture given by S. Havlin in the June 2010 Varenna School. Modern systems tend to be more coupled together. Infrastructures, biology, sociology and economy systems are interconnected such that events taking place in one system may propagate and influence other coupled networks. Recent studies \cite{Buldyrev10,Parshani10} show that coupling between systems increases their vulnerability to random failure. Properties of interdependent networks significantly differ from the ones of a single network. In this article, these results are reviewed and the main properties are discussed. 
  When a system of two interdependent networks is considered, where nodes in one network have a bidirectional coupling with nodes in the other, the percolation properties are significantly affected.
  Due to coupling, not only the transition threshold is increased but also the order of the transition changes.

  The presence of interdependency between nodes in different networks, such that if one of the nodes is inactive the other can not function as well, leads to catastrophic effects when some nodes are removed from the system.
  A cascade of events is then ignited leading to an abrupt decomposition of the mutually connected giant component. 
  For two interconnected ER graphs, when nodes are removed randomly, a percolation transition is observed.
  While for a single network the transition is always second order, for the coupled system the transition is rather first order and the threshold corresponds to much less removed nodes.
  This increase in vulnerability with the coupling is also observed for scale-free networks and, unlike single networks, even for values of the degree exponent below three a percolation transition is observed.

  Tuning the fraction of interdependent nodes shifts the percolation threshold.
  The stronger the coupling the lower the fraction of nodes that needs to be removed to fully fragment the giant component.
  Yet, the order of the transition changes from second order, in the weak coupling regime, to first order under strong coupling.
  In the two-parameter diagram, of coupling and fraction of removed nodes, there are two transition lines, one of first order and the other a second order line that mutually touch in a critical point.

  These interesting results raise new questions in the field of complex networks.
  As a natural follow up, it is interesting to understand what happens when different types of networks are coupled, with special focus on real networks.
  The effect of different types of inter-networks connections is also relevant.
  Besides, understanding how rewiring of the system may improve the resilience to failures is of paramount interest.

  \newpage
  \bibliography{havlin}

\begin{thebibliography}{10}

\bibitem{Albert02}
R.~Albert and \mbox{A.-L.} Barab\'asi.
\newblock Statistical mechanics of complex networks.
\newblock {\em Rev. Mod. Phys.}, 74:47, 2002.

\bibitem{Dorogovstev02}
S.~N. Dorogovtsev and J.~F.~F. Mendes.
\newblock Evolution of networks.
\newblock {\em Adv. in Phys.}, 51:1079, 2002.

\bibitem{Newman03}
M.~E.~J. Newman.
\newblock The structure and function of complex networks.
\newblock {\em SIAM Rev.}, 45:167, 2003.

\bibitem{Newman10}
M.~E.~J. Newman.
\newblock {\em Networks: An Introduction}.
\newblock Oxford University Press, Oxford, 2010.

\bibitem{Dorogovstev08}
S.~N. Dorogovtsev, A.~V. Goltsev, and J.~F.~F. Mendes.
\newblock Critical phenomena in complex networks.
\newblock {\em Rev. Mod. Phys.}, 80:1275, 2008.

\bibitem{Cohen10}
R.~Cohen and S.~Havlin.
\newblock {\em Complex networks: structure, robustness and function}.
\newblock Cambridge University Press, Cambridge, 2010.

\bibitem{Clauset09}
A.~Clauset, C.~R. Shalizi, and M.~E.~J. Newman.
\newblock Power-law distributions in empirical data.
\newblock {\em SIAM Rev.}, 51:661, 2009.

\bibitem{Albert00}
R.~Albert, Jeong H., and A.-L. Barab\'asi.
\newblock Error and attack tolerance of complex networks.
\newblock {\em Nature}, 406:378, 2000.

\bibitem{Cohen01}
R.~Cohen, K.~Erez, D.~\mbox{ben-Avraham}, and S.~Havlin.
\newblock Breakdown of the internet under intentional attack.
\newblock {\em Phys. Rev. Lett.}, 86:3682, 2001.

\bibitem{Albert04}
R.~Albert, I.~Albert, and G.~L. Nakarado.
\newblock Structural vulnerability of the \mbox{North American} power grid.
\newblock {\em Phys. Rev. E}, 69:025103, 2004.

\bibitem{Moreira09}
A.~A. Moreira, J.~S.~Andrade Jr, H.~J. Herrmann, and J.~O. Indekeu.
\newblock How to make a fragil network robust and vice versa.
\newblock {\em Phys. Rev. Lett.}, 102:018701, 2009.

\bibitem{Holme02}
P.~Holme, B.~J. Kim, C.~N. Yoon, and S.~K. Han.
\newblock Attack vulnerability of complex networks.
\newblock {\em Phys. Rev. E}, 65:056109, 2002.

\bibitem{Schneider10}
C.~M. Schneider, A.~A. Moreira, J.~S.~Andrade Jr., S.~Havlin, and H.~J.
  Herrmann.
\newblock Mitigation of malicious attacks on networks.
\newblock {\em Preprint}, 2010.

\bibitem{Herrmann10}
H.~J. Herrmann, C.~M. Schneider, A.~A. Moreira, J.~S.~Andrade Jr., and
  S.~Havlin.
\newblock Onion-like network topology enhances robustness against malicious
  attacks.
\newblock {\em accepted for JSTAT}, 2010.

\bibitem{Buldyrev10}
S.~V. Buldyrev, R.~Parshani, G.~Paul, H.~E. Stanley, and S.~Havlin.
\newblock Catastrophic cascade of failures in interdependent networks.
\newblock {\em Nature}, 464:1025, 2010.

\bibitem{Parshani10}
R.~Parshani, S.~V. Buldyrev, and S.~Havlin.
\newblock Interdependent networks: reducing the coupling strength leads to a
  change from a first to second order percolation transition.
\newblock {\em Phys. Rev. Lett.}, 105:048701, 2010.

\bibitem{Peerenboom01}
J.~Peerenboom, R.~Fischer, and R.~Whitfield.
\newblock Recovering from disruptions of interdependent critical
  infrastructures.
\newblock {\em Pro. CRIS/DRM/IIIT/NSF Workshop Mitigat. Vulnerab. Crit.
  Infrastruct. Catastr. Failures}, 2001.

\bibitem{Rosato08}
V.~Rosato, L.~Issacharoff, F.~Tiriticco, S.~Meloni, S.~De Porcellinis, and
  R.~Setola.
\newblock Modelling interdependent infrastructures using interacting dynamical
  models.
\newblock {\em Int. J. Crit. Infrastruct.}, 4:63, 2008.

\bibitem{Stauffer94}
D.~Stauffer and A.~Aharony.
\newblock {\em Introduction to Percolation Theory}.
\newblock Taylor \& Francis, London, 1994.

\bibitem{Bunde96}
A.~Bunde and S.~Havlin, editors.
\newblock {\em Fractals and Disordered Systems}.
\newblock Springer, Heidelberg, 1996.

\bibitem{Erdos59}
P.~Erd\H{o}s and A.~R\'enyi.
\newblock On random graphs. \mbox{I}.
\newblock {\em Publ. Math. (Debrecen)}, 6:290, 1959.

\bibitem{Erdos60}
P.~Erd\H{o}s and A.~R\'enyi.
\newblock On the evolution of random graphs.
\newblock {\em Publ. Math. Inst. Hung. Acad. Sci.}, 5:17, 1960.

\bibitem{Bollobas85}
B.~Bollob\'as.
\newblock {\em Random Graphs}.
\newblock Academic Press, London, 1985.

\bibitem{Barabasi99}
A.-L. Barab\'asi and R.~Albert.
\newblock Emergence of scaling in random networks.
\newblock {\em Science}, 286:509, 1999.

\bibitem{Cohen02}
R.~Cohen, D.~\mbox{ben-Avraham}, and S.~Havlin.
\newblock Percolation critical exponents in scale-free networks.
\newblock {\em Phys. Rev. E}, 66:036113, 2002.

\bibitem{Cohen00}
R.~Cohen, K.~Erez, D.~\mbox{ben-Avraham}, and S.~Havlin.
\newblock Resilience of the internet to random breakdowns.
\newblock {\em Phys. Rev. Lett.}, 85:4626, 2000.

\bibitem{Achlioptas09}
D.~Achlioptas, R.~M. D'Souza, and J.~Spencer.
\newblock Explosive percolation in random networks.
\newblock {\em Science}, 323:1453, 2009.

\bibitem{Ziff09}
R.~M. Ziff.
\newblock Explosive growth in biased dynamic percolation on two-dimensional
  regular lattice networks.
\newblock {\em Phys. Rev. Lett.}, 103:045701, 2009.

\bibitem{Radicchi09}
F.~Radicchi and S.~Fortunato.
\newblock Explosive percolation in scale-free networks.
\newblock {\em Phys. Rev. Lett.}, 103:168701, 2009.

\bibitem{Radicchi10}
F.~Radicchi and S.~Fortunato.
\newblock Explosive percolation: A numerical analysis.
\newblock {\em Phys. Rev. E}, 81:036110, 2010.

\bibitem{Araujo10}
N.~A.~M. Ara\'ujo and H.~J. Herrmann.
\newblock Explosive percolation via control of the largest cluster.
\newblock {\em Phys. Rev. Lett.}, 105:035701, 2010.

\bibitem{DaCosta10}
R.~A. da~Costa, S.~N. Dorogovtsev, A.~V. Goltsev, and J.~F.~F. Mendes.
\newblock \mbox{"Explosive} percolation" transition is actually continuous.
\newblock {\em arXiv:1009.2534v2}, 2010.

\bibitem{Newman02}
M.~E.~J. Newman.
\newblock Spread of epidemic disease on networks.
\newblock {\em Phys. Rev. E}, 66:016128, 2002.

\bibitem{Shao08}
J.~Shao, S.~V. Buldyrev, R.~Cohen, M.~Kitsak, S.~Havlin, and H.~E. Stanley.
\newblock Fractal boundaries of complex networks.
\newblock {\em Europhys. Lett.}, 84:48004, 2008.

\bibitem{Shao09}
J.~Shao, S.~V. Buldyrev, L.~A. Braunstein, S.~Havlin, and H.~E. Stanley.
\newblock Structure of shells in complex networks.
\newblock {\em Phys. Rev. E}, 80:036105, 2009.

\end{thebibliography}


\end{document}